\documentclass[twocolumn,nofootinbib]{revtex4}%

\def\ben{\begin{equation}}
\def\een{\end{equation}}
\def\bena{\begin{eqnarray}}
\def\eena{\end{eqnarray}}

\newcommand{\be}{\begin{equation}}
\newcommand{\ee}{\end{equation}}
\newcommand{\bea}{\begin{eqnarray}}
\newcommand{\eea}{\end{eqnarray}}
\newcommand{\beq}{\begin{eqnarray}}
\newcommand{\eeq}{\end{eqnarray}}

\def\6{\partial}

\def\a{\alpha}

\def\n{\nu}

\def\B{\begin{equation}}
\def\E{\end{equation}}

\usepackage{amssymb}
\usepackage{amsfonts}
\usepackage{amsmath}
\usepackage{graphicx}

\makeatletter

\@addtoreset{equation}{section}
\makeatother

\begin{document}
\title{Extended Charged  Events and Chern-Simons Couplings}
\author{ Claudio Bunster$^{1}$, Andr\'es Gomberoff$^{2}$ and Marc Henneaux$^{1,3}$}

\affiliation{$^{1}$Centro de Estudios Cient\'{\i}ficos (CECs), Casilla 1469, Valdivia, Chile}
\affiliation{$^{2}$Universidad Andres Bello, Departamento de Ciencias F\'{\i}sicas, Av. Rep\'ublica 252, Santiago, Chile.}
\affiliation{$^{3}$Universit\'e Libre de Bruxelles and International Solvay Institutes, ULB-Campus Plaine CP231, B-1050 Brussels, Belgium}

\begin{abstract}
In three spacetime dimensions the worldvolume of a magnetic source is a single point, a magnetically charged event. It has been shown long ago that in three-dimensional spacetime the Chern-Simons coupling is quantized, because the magnetic event emits an electric charge which must be quantized according to the standard Dirac rule. Recently, the concept of dynamical extended charged events has been introduced, and it has been argued that they should play as central a role  as that played by particles or ordinary branes. In this article we show that in the presence of a Chern-Simons coupling, a magnetically charged extended event emits an  extended object, which geometrically is just like a Dirac string, but it is observable, obeys equations of motion, and may be electrically charged. We write a complete action principle which accounts for this effect. The action involves two Chern-Simons terms, one integrated over spacetime and the other integrated over the  worldvolume of the submanifold that is the union of the Dirac world-sheet and the history of the emitted physical object.  By demanding  that the total charge emitted by a composite extended magnetic event be quantized according to Dirac's rule, we find a quantization condition for the Chern-Simons coupling.  For a $1$-form electric potential in $D=2n+1$ spacetime dimensions, the composite event is formed by $n$ elementary extended magnetic events separated in time  such that the product of their transverse  spaces, together with the time axis, is the entire spacetime. We show that the emitted electric charge is given by the integral  of the $(n-1)$-th exterior power of the electromagnetic field strength over the last elementary event, or, equivalently,  over an appropriate closed surface. The extension to more general $p$-form potentials and higher dimensions is also discussed. For the case $D=11$, $p=3$, our result for the quantization of the Chern-Simons coupling was obtained previously in the context of M-theory,  a remarkable agreement that makes the existence of events even more compelling.  
\end{abstract}

\pacs{11.15.-q,11.15.Yc,14.80.Hv}
\maketitle

\section{Introduction}
\setcounter{equation}{0}
Dynamical charged events where introduced in \cite{Bachas:2009ve}. It was argued there that they are forced upon us by  the principle of electric-magnetic duality, which states that electric and magnetic fields must be treated on equal footing, and that therefore one must consider both electric and magnetic sources in the Maxwell equations. Unlike ordinary particles or branes, events obey no equations of motion within the spacetime where they happen. However, they do produce fields within that spacetime, acting as sources according to the standard field equations. These fields act, in turn, on ordinary charged particles and branes. In order to have a closed system in which events obey the principle of action and reaction, rather than being just external sources, one must endow them with dynamics. This may be done by considering ordinary spacetime as embedded in a larger one, and describing events as the imprint left on the embbeded spacetime by an object impinging on it from an extra dimension. Thus, a charged point event is considered as the imprint of a particle which hits spacetime. Similarly for extended events.

A point event is the simplest member of a family which includes also {\it extended events}. An extended event is the imprint left on spacetime when a $p$-brane, with $p\ge 1$, hits it. Just as point events, extended events do not obey an equation of motion in the $D$-dimensional spacetime because their dynamics takes place in the extra dimensions. This is the key difference between events and particles, or ordinary extended objects. For example, a closed one-dimensional event is a loop in the  $D$-dimensional spacetime with arbitrary shape: it can be timelike, spacelike, null, or a combination of all three. It can also go forward and backwards in time. The observation of an event would not be a violation of causality or of any conservation law, it would rather be evidence for the presence of extra dimensions. 

In contradistinction with events, ordinary particles, or branes, with either electric or magnetic charge, live within the $D$-dimensional spacetime, and obey action principles and ordinary dynamical laws there.

It should be stressed that events occur in Lorentzian spacetime, and are to be distinguished from instantons, which are solutions of the Euclidean equations and describe quantum mechanical tunneling. Also the fact that events, when viewed as a result of an object impinging from an extra dimension, need no fine tuning, contrasts with the point of view adopted in the ``S-brane" literature, as advocated, for instance, in \cite{Durin:2005ts}.

The  previous remarks have been made to situate events within the theoretical framework. In the present article we shall employ particular configurations of extended events and discuss their influence on ordinary branes and fields, according to an action principle which involves only the $D$-dimensional spacetime.  We will therefore make no use here of the fact that the charged events are imprints from extra dimensions, other than occasional conceptual comments.

The plan of the paper is the following. Sec. \ref{Dirac} recalls briefly the standard theory of magnetic poles and their associated strings, given by Dirac \cite{Dirac:1948um} in 1948, and its generalization to extended objects developed in \cite{Teitelboim:1985yc}. It is observed that the fundamental property that a Dirac string cannot go through a charged particle, the ``Dirac veto", holds as well for Dirac strings emanating from magnetic events and that, as a consequence, the standard quantization rule for the product of electric and magnetic charges remains valid. 

Next, in Sec. \ref{chernsimons},  the treatment is extended to the simplest non-linear Chern-Simons coupling,  a $1$-form electric potential in $D=5$ dimensions. An action principle is given. Its most striking new feature is that it brings in a new object into the theory. It is an extended object emitted by the magnetic event, whose   history is a semi-infinite surface, which geometrically is just like the worlsheet of the Dirac string, but it is observable, obeys equations of motion, and may be electrically charged.
The Chern-Simons action (\ref{CS})  is then the sum of {\it two Chern-Simons terms}: the standard one for spacetime, and an aditional one over an embedded surface $\Sigma$ which contains the magnetic event, and it is the union of the string worldsheet and the history of the physical emitted object.

It is found that, due to the non-linearity,  the Dirac veto takes a stronger form:  As before, Dirac strings cannot cross electric worldlines, but now they cannot cross among themselves either. This applies both to Dirac strings emanating from different magnetic events (or poles) as well as to self intersections of one Dirac string (The strengthening of the Dirac veto was previously established, in a different setting, in Ref. \cite{Bekaert:2002eq}). 

Next, in Sec. IV, we demand that the electric charge emitted by the magnetic event, which is proportional to the Chern-Simons coupling, should be quantized according to the standard Dirac rule, and thus infer the quantization of that coupling, generalizing the result of Ref. \cite{Henneaux:1986tt} for $D=3$. To that effect we introduce a composite event formed by two extended magnetic events separated in time, and such that the product of their transverse  spaces, together with the time axis, is the entire spacetime. We obtain directly in this way, a result  previously arrived at using dimensional reduction and the Witten effect \cite{Witten:1979ey} in four spacetime dimensions \cite{Bachas:1998rg}. 

Finally, in that section, we show that the emitted electric charge is given by the first Chern class of the electromagnetic field over a two-dimensional surface linking the first magnetic event.  

Next in Sec. V we extend the analysis to higher $D$ and $p$. For a $1$-form electric potential in $D=2n+1$ spacetime dimensions, the composite object is formed by $n$ extended magnetic events separated in time  such that the product of their transverse  spaces, together with the time axis, is the entire spacetime. We pay special attention to the $3$-form appearing in supergravity.  Sec. VI is devoted to concluding remarks.

\section{Dirac strings and Dirac veto}
\label{Dirac}

\subsection{Pure Maxwell theory}
In his classical paper \cite{Dirac:1948um} of 1948, Dirac invented an action principle for the electromagnetic field interacting with both electric and magnetic sources. In what we would now call the ``electric representation", the $1$-form electric potential $A_\mu$ couples minimally to electric sources, in the standard way. On the other hand, the coupling to the magnetic sources is achieved through the introduction of a Dirac string which emanates from the magnetic pole. In the case of  ordinary point particles carrying magnetic charge studied by Dirac, the string is indeed a string, that is, a semi-infinite one dimensional line whose boundary is the magnetic pole. As the magnetic pole traces a one dimensional worldline the Dirac string traces a two dimensional worldsheet whose boundary is the magnetic worldline. If the embedding of the  worldsheet is described by 
\begin{equation}
\label{yy}
x^\mu=y^\mu(\xi^1, \xi^2) , \ \ \ \    -\infty<\xi^1<\infty, \ \  0<\xi^2<\infty
\end{equation}
then the worldline of the magnetic pole is
\begin{equation}
\label{zz}
x^\mu=z^\mu(\xi_1)=y^\mu(\xi^1, \xi^2=0) .
\end{equation}
We will denote the string worldsheet  (\ref{yy}) by $\Sigma_{+}^2$, where the subindex $2$ indicates its dimension and the superscript $+$ recalls that $\xi^2>0$.

The coupling of the magnetic source to the electromagnetic field is achieved by modifying the definition of the electromagnetic field strength
\begin{equation}
F=dA
\end{equation}
to read instead,
\begin{equation}
\label{G}
F=dA + {}^* G_{+},
\end{equation}
where
\begin{equation}\label{GG}
G^{\mu\nu}_+=g\int_{\Sigma^2_+} \delta^{(4)}(x-y)dy^\mu\wedge dy^\nu .
\end{equation}
The two-form $G_+$ is the current associated to the string worldsheet.  The current $G_+$ is not conserved because $\Sigma^2_+$ has a boundary on the magnetic pole. One has 
\begin{equation}
\label{}
\partial_\mu G^{\mu\nu}_+ (x) = j^\nu_{mag} (x) = g\int \delta^{(4)} (x-z)dz^\nu.
\end{equation}

The Maxwell action reads
\begin{equation}
I_M=-\frac{1}{4}\int F^{\mu\nu}F_{\mu\nu} d^4 x ,
\end{equation}
with $F_{\mu\nu}$ given by (\ref{G}). It depends on both $A_\mu(x)$ and $y^\mu(\tau, \sigma)$. If one varies $I_M$ with respect to the worldline $z^\mu(\tau)$ of the magnetic source one finds,
\begin{equation}
\delta I_M= g\int_{\mbox{ \tiny worldsheet}}  (d{}^* F)_{\mu\nu\rho}\delta y^\rho dy^\mu\wedge dy^\nu  ,
\end{equation}
which on account of the Maxwell equation, 
\begin{equation}
\label{maxwell}
d{}^* F={}^*j_{el},
\end{equation}
becomes,
\begin{equation}
\label{dIM}
\delta I_M= g\int_{\mbox{ \tiny worldsheet}}  ({}^*j_{el})_{\mu\nu\rho}\delta y^\rho dy^\mu\wedge dy^\nu .
\end{equation}
The variation (\ref{dIM}) vanishes automatically provided that the string worldsheet is never crossed by an electric worldline. This is the original Dirac veto. Note that  the veto allows $j^{\mu}$ to be tangent to the string worldsheet because then the integrand in Eq. (\ref{dIM}) vanishes. No restriction is imposed on the shape of the worldsheet itself, which is a reflection of the fact that the $y^\mu(\xi)$ may be thought of as purely gauge variables  for $\xi^2>0$. For $\xi^2=0$, Eq. (\ref{zz}) must hold.

When one passes to quantum mechanics, the Dirac veto, combined with the pure gauge character of the string, leads to the quantization condition \cite{Dirac:1931kp, Dirac:1948um} for the product of the electric charge $e$ and the magnetic charge $g$, 
\begin{equation}
\label{eg}
eg=2\pi \hbar m ,
\end{equation}
where $m$ is an integer. That analysis will not be repeated here.

\subsection{Extended Sources}

The preceding discussion can be generalized to deal with electric and magnetic extended sources coupled to a $p$-form electric potential. The analysis, the formulae and the conclusions apply equally well to the case of  ordinary branes and extended events. 
If the spacetime is $D$ and the dimension of the ``worldline" of the electric extended object is $p$, then the dimension of the worldline of the magnetic electric object is $D-p-2$. The string ``worldsheet" has $D-p-1$ dimensions, and its boundary is the magnetic worldline. 

Eq. (\ref{y}) now reads,
\begin{equation}
\label{y}
x^\mu=y^\mu(\xi^1,\ldots \xi^{D-p-1}) , 
\end{equation}
with, 
\begin{equation}
\label{range}
\xi^{D-p-1}\ge 0,
\end{equation}
while Eq. (\ref{z}) becomes,
\begin{eqnarray}
\label{z}
x^\mu&=&z^\mu(\xi^1,\ldots \xi^{D-p-2}) \nonumber\\
&=&y^\mu(\xi^1,\ldots \xi^{D-p-1}, 0) .
\end{eqnarray}

The $(p+1)$-form field strength $F$ now reads, 
\begin{equation}
\label{F=dA+G}
F=dA + (-1)^{(p+1)(D+1)} {}^* G_+,
\end{equation}
but now one has
\begin{equation}
G_+^{\mu_1\ldots \mu_{D-p-1}}=g\int_{\Sigma_+^{D-p-1}} \hspace{ -0.5cm} \delta^{(D)}(x-y)dy^{\mu_1}\wedge\cdots \wedge dy^{\mu_{D-p-1}} .
\end{equation}

The Maxwell action now reads,
\begin{equation}
I_M=-\frac{1}{2}\frac{1}{(p+1)!}\int F^{\mu_1\ldots \mu_{p+1}}F_{\mu_1\ldots \mu_{p+1}} d^D x .
\end{equation}
Its variation with respect to the string coordinates is given by, 
\begin{equation}
\label{dIMg}
\delta I_M= g\int  ({}^*j_{el})_{\mu_1\cdots \mu_{D-p-1} \rho} \delta y^\rho dy^{\mu_1} \wedge \cdots \wedge dy^{\mu_{D-p-1}} ,
\end{equation}
and demanding that it vanishes is again equivalent to imposing the Dirac veto. 

In this more general case the electric and magnetic charges are not dimensionless. Rather, $e$ has units of length to the power of $(D-2p-2)/2$ . However, the product of $e$ and $g$ is still dimensionless and the quantization condition (\ref{eg}) continues to hold \cite{Teitelboim:1985yc}.

\section{Addition of a Chern-Simons coupling for $D=5$ and $p=1$}
\label{chernsimons}

\subsection{Action}

We will first postulate an action and discuss its main general properties and implications. Next we will derive the equations of motion and discuss the phenomenon of emission of electric charge by a magnetic source.

The most striking new feature of the introduction of a Chern-Simons coupling is that, in order to write the action principle, one needs to bring in a new geometrical object into the theory. It is a semi-infinite surface, which, for $D=5$ has three dimensions. 

The boundary of this new surface is the magnetic event, just as it is the case with the Dirac worldsheet $\Sigma^3_+$. But unlike the latter, the new surface is the history of a two-dimensional physical extended object, which obeys equations of motion. This new object, which emerges from the magnetic event, may have electric charge.  
We will call its worldsheet $\Sigma^3_-$.  

The equations of its embedding in the $D=5$ spacetime may be written just as those of the Dirac string, but with the parameter $\xi^3$ taking negative values. 
We will denote plainly by $\Sigma^3$, the glueing, at their common boundary $\xi^3=0$ (the magnetic event), of $\Sigma^3_+$, and $\Sigma^3$, as illustrated in Fig 1. The surface $\Sigma^3$, is infinite in all directions, and its associated current:
\begin{equation}\label{GGG}
G^{\mu\nu\rho}(x)=g\int_{\Sigma^3} \delta^{(5)}(x-y)dy^\mu\wedge dy^\nu\wedge dy^\rho ,
\end{equation}
is conserved,
\begin{equation}
\label{}
\partial_\mu G^{\mu\nu\rho}= 0.
\end{equation}
The Chern-Simons action (\ref{CS})  will be the sum of {\it two Chern-Simons terms}: the standard one for a five dimensional spacetime, and an aditional one over an embedded three dimensional surface $\Sigma^3$ which contains the magnetic event,
 \begin{equation}\label{CS}
I_{CS}= \frac{\alpha}{6}\int dA\wedge dA \wedge A + \frac{\alpha g}{2}\int_{\Sigma^3} dA \wedge A .
\end{equation}

Note that normally one would couple a three dimensional worldline to a fundamental $3$-form $A_{\mu\nu\rho}$. Then the associated charge would not be the electric charge, but rather, the charge associated with that fundamental $3$-form (see, for example, Ref. \cite{Teitelboim:1985ya}). However, here, the three form is not fundamental since it is constructed out of the $1$-form $A$ as the Chern-Simons $3$-form  $ dA\wedge A$. As a consequence, we will find below that, generically, there is a contribution to the electric current which has support on the worldsheet $\Sigma_-$, in addition to the one coming from the ordinary minimal coupling to worldlines.  In that sense, the extended object whose history is $\Sigma_-$ may be electrically charged.

One is in the presence of an interesting sequence. Particles whose worldline has dimension one couple to the one dimensional Chern-Simons term $A$. Two dimensional extended objects, whose worldline has dimension three, couple through the Chern-Simons term of dimension three,  $ dA\wedge A$.  The possibility of coupling extended objects to a $1$-form through the composite Chern-Simons form built from it, has been emphasized in Refs. \cite{Zanelli:2008sn},  \cite{Edelstein:2010sh}.

\begin{figure}[h]
\label{fig} 
\begin{center}
  \includegraphics[width=8cm]{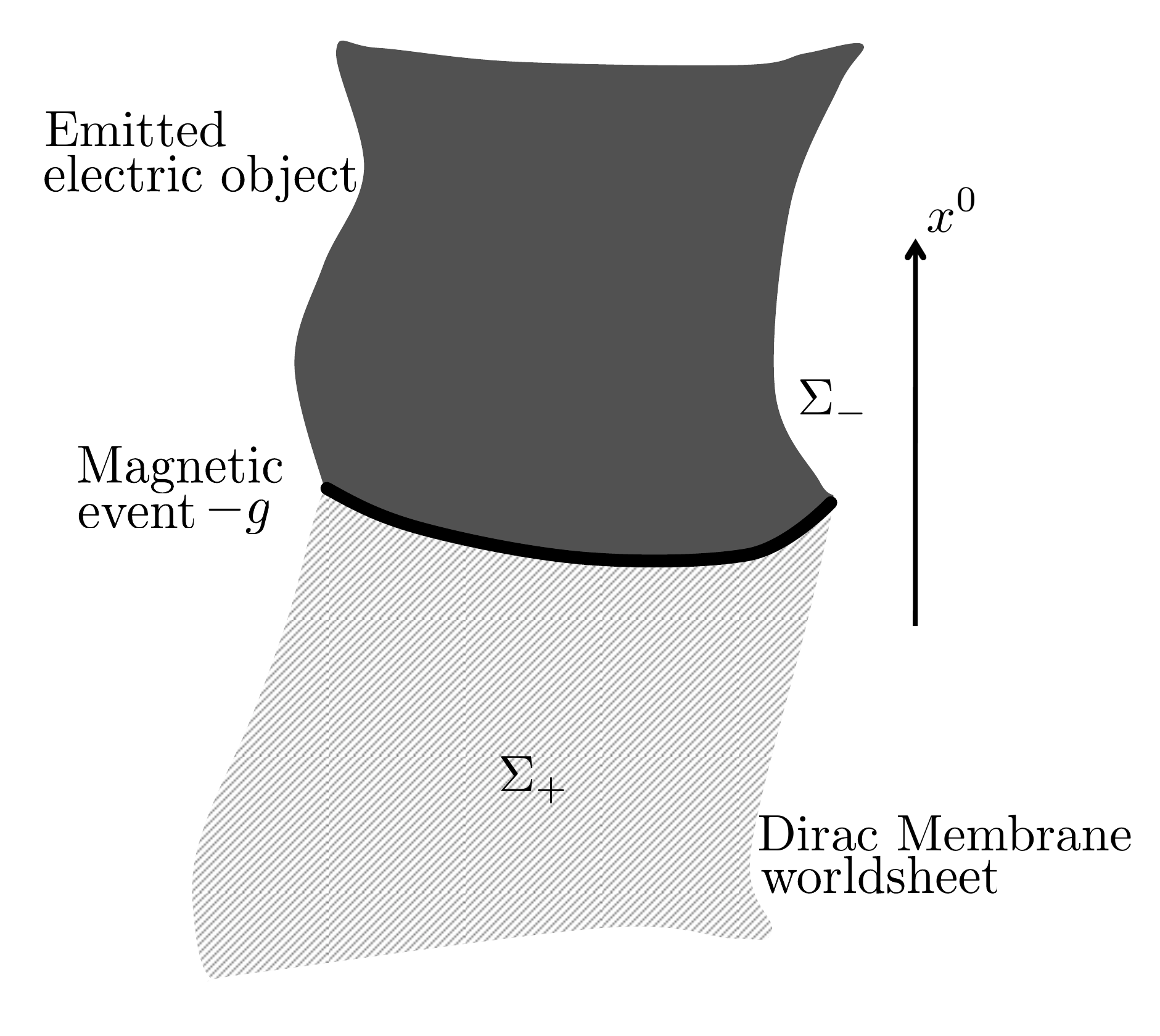}
  \caption{The surface $\Sigma$ for a single magnetic event. 
  The magnetic event  of strength $-g$ is the common boundary of two different  worldvolumes: the Dirac string worldsheet (shaded), parametrized with $\xi^3>0$ and the extended electric object worldsheet (dark), with $\xi^3<0$. The Dirac string has an associated current $G^{\alpha\beta\gamma}_+$ supported on it,  while the extended electric object has $G^{\alpha\beta\gamma}_-$. The complete surface is infinite in all directions and has an associated conserved current  $G^{\alpha\beta\gamma}=G^{\alpha\beta\gamma}_+  + G^{\alpha\beta\gamma}_-$.  We have chosen $-g$ as the charge of the magnetic event in order that, with the conventions used in the main text, the electric object is emitted to the future.  Then  $\partial x^0/\partial \xi^3 <0$.}
   \end{center}
\end{figure}

The complete action  will be the sum of four terms, 
\begin{equation}\label{act}
I=I_M+I_{CS} -\int {}^*j_{particles}\wedge A + I_{K}.
\end{equation}
Here $j_{particles}$ is the conserved electric current of a collection of point particles,   
$$
j^{\mu}_{particles} = \sum q_{i} \int\delta^{(5)}(x-w_{(i)}) dw^{\mu}_{i} ,
$$
where $q_i$ and  $w_{i}^\mu(\tau)$ are the electric charge and worldline of the $i$-th electric charge.
The kinetic term $I_K$ is the sum of two terms, one for the electrically charged point particles, and another for the new object with a three dimensional worldline that we just described,
\begin{equation}\label{}
I_K=I_{particles} + I_-.
\end{equation}
For each charged particle, the kinetic action is proportional to its worldlength, 
\begin{equation}\label{kinetic}
I_{particles} =  -\sum_i m_{i} \int \sqrt{-\dot{w}_{i}^\mu \dot{w}_{i\mu}}\ d\tau,
\end{equation}
whereas for the electrically charged extended object the action is proportional to its worldvolume,
 \begin{equation}\label{}
I_- = -M\int_{\Sigma^3_-} \sqrt{-{}^{(3)}\gamma} \ d\xi^1 d\xi^2 d\xi^3 ,
\end{equation}
Here ${}^{(3)}\gamma$ is the determinant of the induced metric on the worldsheet
$\gamma_{rs} = \partial_r y^{\mu} \partial_s y_{\mu}$, with $\partial_r=\partial/ \partial \xi^r$.

\subsection{Equations of motion and Dirac veto}

 We will now proceed to demand that the action (\ref{act}) be stationary with respect to variations of the potential $A$, of the particle worldlines $w_i(\tau)$, and of the complete three dimensional worldsheet $y(\xi)$, which includes the Dirac string of the magnetic event and the newly introduced dynamical extended object.
 
 Upon variation of the potential $A$, we find
 \begin{equation}
\label{mcs}
d{}^* F + \frac{1}{2}\alpha F\wedge F={}^*j_{particles} + {}^*j_- ,
\end{equation}
with,
\begin{equation}\label{jm}
j^\mu_-(x) = \alpha g \int_{\Sigma^3_-}\delta^{(5)}(x-y) F\wedge dy^\mu
\end{equation}
We also recall that it follows from the definition (\ref{G}) of $F$ that
\begin{equation}\label{monopoles}
dF={}^*j_{mag}
\end{equation}
Next, upon variation of the  $w^\mu_i(\tau)$, we find that each charged particle couples to the $1$-form $A$ through the Lorenz force law,
 \begin{equation}\label{}
m_i\ddot{w_i}^\mu = q_i F^{\mu}_{\ \nu }\dot{w}^\nu_i,
 \end{equation}
whereas, upon variation of the $y^\mu_i(\xi)$ for $\xi^3<0$, we find that  the extended charged object couples to $A$ according to,
\begin{equation}\label{eo}
M\Box y^\mu(\xi) =(\alpha g)\epsilon^{rst} (-\gamma)^{-1/2} F^{\mu}_{\  r} F_{st}
\end{equation}
Now we compute the variation of the action with respect to  $y^\mu_i(\xi)$ for $\xi^3>0$ (Dirac string). The variation of the Dirac string, $\delta y$ may be defined choosing a  four dimensional manifold whose boundary is the union of  the original and the varied Dirac world-volumes,  $y^\mu$ and  $y^\mu + \delta  y^\mu$. This manifold has an associated current $\Gamma^{\alpha\beta\gamma\delta}$, satisfying,
\begin{equation}\label{}
d{}^* \Gamma=\delta {}^*G_+
\end{equation}
We use this last expression to write the variation of the action with respect to the Dirac string,
\begin{equation}\label{veto}
\delta I = \int \left(    d{}^*F +\frac{\alpha}{2}dA\wedge dA \right)\wedge  {}^*\Gamma ,
\end{equation}
Making use of Eq. (\ref{mcs}) we see that  the variation (\ref{veto}) vanish automatically provided that
 (i) the three dimensional worldsheet of the Dirac string cannot be crossed by an electric current, (ii) the Dirac string should not intersect with itself. If there are several magnetic sources one finds, in the same way, that their Dirac strings cannot intersect with each other. 
 This is the  Dirac veto, which due to the presence of the Chern-Simons coupling takes a stronger form than for the pure Maxwell case.
 This strengthening of the Dirac veto does not yield any new quantization condition but, without it, Eqs. (\ref{mcs})-(\ref{eo}) do not follow from an action principle.
 
 In this context, it is of interest to observe that an important property already found in Ref. \cite{Henneaux:1986tt} for $D=á3$, also holds here.
The property is the following: although, at any given time, the individual electric charge of each charged object (point or extended)  is well defined, and therefore so it is the total charge, there is no unambiguous notion of charge density in space. This is due to the fact that the density of electric charge appearing on the right side of the Gauss law is not the one on the right side of (\ref{mcs}), but rather, the one obtained by replacing $j_-^\mu$ in that equation by
\begin{equation}
\label{jj}
j^\mu(x) = \alpha g \int_{\Sigma^3}\delta^{(5)}(x-y) dA\wedge dy^\mu.
\end{equation}
Now, this $j^\mu$ is conserved and invariant under $A\rightarrow A+ d\Lambda$, but, unlike $j^\mu$ ,  it changes if one displaces the string worldsheet piece of it coming from $\Sigma^3_+$. For example, one may make $\Sigma^3_+$ to cross any $x^0=$constant surface back and forth in time, thus changing the amount of charge present within any space volume. On the other hand, the charges of the emitted object are unambiguous  because of the Dirac veto, which forbids putting a Dirac string on top of them. See in this context also \cite{Marolf:2000cb}.

\subsection{Emission of electric charge by a magnetic source} \label{A2}
If we look at the right hand side of  (\ref{mcs}) we observe that the current $j^\mu_-$ given by (\ref{jm})
appears on the same footing with the electric current of the particles. Unlike the current of the particles, the current $j_-^\mu$ has its support on the semi-infinite three dimensional surface $\Sigma^3_-$, and, more importantly, it is not conserved. Indeed, by direct calculation from  (\ref{jm}) one finds
\be 
\label{nc}
\partial_\n j^\n_{-} (x)=- \alpha g  \int \delta^{(5)}(x-z)F(z)
\ee 
where $z(\xi^1, \xi^2)$ is the magnetic event.

Eq.  (\ref{nc}) makes the exterior derivative of Eq  (\ref{mcs}) to be an identity. This consistency requirement is actually what motivated the introduction in the action of the complete $\Sigma^3$, including $\Sigma^3_-$ , in addition to the Dirac string  $\Sigma^3_+$ required by the magnetic event.

If one has several magnetic events $z_i$ with magnetic charges $g_i$, one replaces Eq.  (\ref{nc}) by 
\be 
\label{nc2}
\partial_\n j^\n_{-} (x)=- \alpha \sum_i g_i  \int \delta^{(5)}(x-z_i)F(z_i)
\ee 

If we take the integral of  (\ref{nc2}) over a five dimensional volume which contains the magnetic event, we obtain that {\it a magnetic event of magnetic charge $g$ emits an electric charge} $Q$ given by  
\be 
\label{Q}
Q_{emitted}=- \alpha g  \int_{event} F(z) \; \; \; \; (D=5).
\ee 

There is an important difference with the case $p=1$, $D=3$. There, the 2-form $F$ is absent in the analog of (\ref{jm}), and therefore the electric charge emitted by a particular magnetic source is totally determined by that source itself, and it is always equal to $-\alpha g$. However, as seen from Eq. (\ref{Q}), for $D=5$ one already sees a more general new effect: the electric charge emitted by a magnetic source depends also on the value of the total electromagnetic field, and therefore on the other sources. The term ``induced emission" would appear appropriate to describe this phenomenon.

A consequence of the induced character of the emission of charge, which, again, is not present for $D=3$ is that the electric current (\ref{jm}) may vanish.  Then the emitted object is still present but it is electrically neutral. This happens most strikingly when the projection $F_{rs}$ of $F$ on $\Sigma^3$ vanishes. In that case there is no driving force on the right hand side of (\ref{eo}) and the emitted object moves as a free two dimensional Nambu string.

We stress again that, as it was pointed out in \cite{Bachas:2009ve},  the present description should be considered as just an effective theory because it cannot properly account for the detailed mechanism of the impact and of the associated energy momentum conservation. Therefore, the boundary conditions are not  determined, and one must make a choice for which solution of the field equation of motion is to be taken. It was argued in  \cite{Bachas:2009ve} that, at least at the classical level,  a natural solution corresponding for a confined flux impinging from the past was the purely retarded one, and we will adhere to that practice below. For the same reason the boundary conditions for the emitted physical object are not fully determined. Its initial position is the magnetic event,
but its initial velocity (or final position) has to be specified. This freedom will be used in Secs. IV and V to choose particular configurations.

\section{Quantization of the Chern-Simons coupling for $D=5$ and $p=1$}

In this section, we will generalize the approach that was used in \cite{Henneaux:1986tt} to obtain the quantization of the Chern-Simons coupling.  There, we evaluated the charge $Q$ emitted by a single event and demanded that it be quantized according to the general rule (\ref{eg}).  The charge turned out to be 
\begin{equation}
Q = - \alpha \, g \hspace{1cm} (D=3), 
\end{equation}
and therefore the quantum of coupling turned out to be 
\begin{equation}
\alpha g^2 = \pi \hbar \hspace{1cm} (D=3).
\end{equation}

We will see below that for $D=5$, the charge emitted by a single magnetic source vanishes and therefore a more elaborate configuration needs to be used in order to derive the quantization rule for $\alpha$. The simplest configuration that we have found is a ``composite event" that consists of two magnetic sources which are complementary and do not cross each other.  One might take one of the sources to be the history of an ordinary extended magnetic pole, but the other one would necessarily have to be a spacelike  extended magnetic event.  Thus, spacelike events are essential to the procedure.  Since one must have at least one of them, the configuration that we will study consists of two spacelike events,  in order to make the computation more symmetric.

\subsection{Field of a single magnetic event}

Take an event A, with magnetic charge $g$, whose world volume is the ($x^3,x^4$)-plane at $x^0= x^1=x^2=0$. For this event, the only non-vanishing component of the magnetic current is 
\begin{equation}
\label{EventA}
j_{A, mag}^{34} = g \delta(x^0) \delta(x^1) \delta(x^2).
\end{equation}
Now, due to the translational and rotational symmetry of the sources in the $(x^3,x^4)$-plane, the problem will be reduced to one in three dimensions ($x^0, x^1,x^2$) with a point source at the origin.  The only non-vanishing components of the field strength $F$ produced by the source (\ref{EventA}), i.e., its self-field,  are given by
\begin{equation}
\label{FA}
F_{\mu \nu}^A = \frac{g}{2 \pi} \frac{\epsilon_{\mu \nu \rho} x^\rho}{\left(-R^2\right)^{\frac{3}{2}}} \theta(-R^2) \theta(x^0)  .
\end{equation}

Here, all the indices are referred to the effective three-dimensional spacetime with coordinates $x^0, x^1, x^2$;   $R^2$ is the invariant  spacetime distance from the effective point source at the origin,
\begin{equation}\label{44}
R^2 = - \left(x^0\right)^2 + \left(x^1\right)^2 + \left(x^2\right)^2,
\end{equation}
and we have taken,   just as in \cite{Bachas:2009ve}, the retarded solution.
The field (\ref{44}) is the retarded field of a pure Maxwell magnetic pole for $D=3$. It turns out to be  an exact  solution of the full Maxwell-Chern-Simons equations (\ref{mcs}) for $D=5$ because it  obeys $F\wedge F=0$ (this does not happen for $D=3$!). 

Since the field  (\ref{44}) is transverse to the magnetic event, it has no projection on it. As discussed above this implies that  the current $j^\mu_-$ in (\ref{jm}) vanishes. This means that the emitted object (which in this case is absorbed, rather) is neutral and moves as a free two dimensional Nambu string.
Note that the argument is purely algebraic based on the symmetry of the problem so that the singularity in the self-field cannot contribute to the integrand in the expression (\ref{Q}) for $Q$.
The fact that the total charge produced by the self-field of the event is zero makes the above configuration not useful to obtain a quantization condition.  

\subsection{Field of a composite magnetic event}

We will now consider a composite two-dimensional event which consists of two complementary spacelike events A, B separated in time.

The event A will be the one already discussed.  The additional event $B$ will have charge $-g$ and it will be located at the complementary  $(x^1,x^2)$-plane at a later time $x^0=a >0$, so that the equations describing event B are $x^0 = a, x^3=0, x^4 = 0$ and the magnetic current is 
\begin{equation}
\label{EventB}
j_{B, mag}^{12} =  -g \delta(x^0 - a) \delta(x^3) \delta(x^4).
\end{equation}

The geometrical configuration in consideration  may be described as follows. An electric object comes in and ends on the magnetic event A.  The magnetic event A,  in turn, emits a Dirac membrane that is next absorbed by event B, which finally proceeds to emit another electric object.  This configuration is  illustrated in Fig. 2. It consists of two dark patches connected by a shaded patch in between.

\begin{figure}[h]
\label{fig55} 
\begin{center}
  \includegraphics[width=8cm]{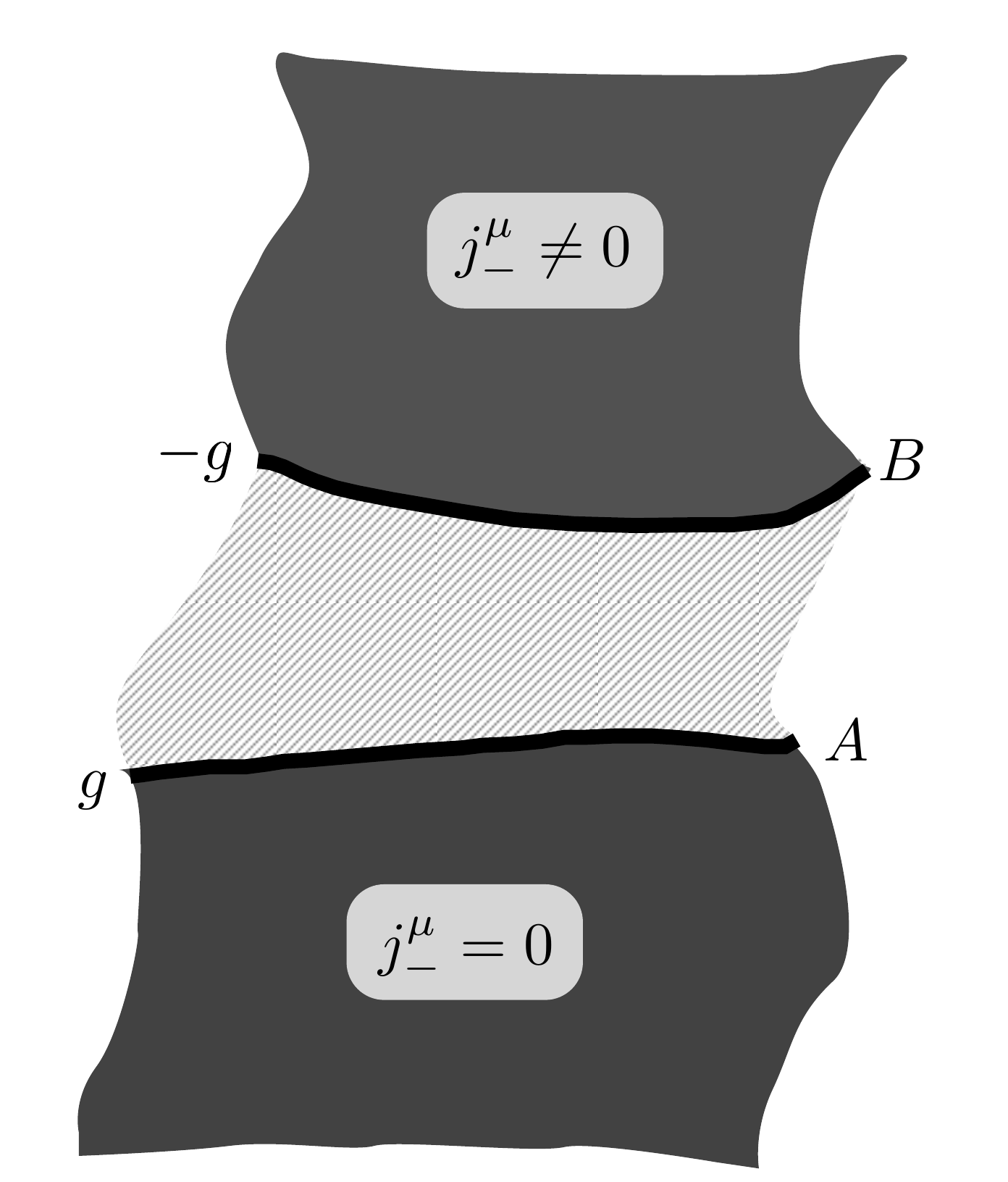}
  \caption{The surface $\Sigma$ for $D=5$. The surface shown in  Fig. 1 contains only one magnetic event. One may extend the notion to a surface $\Sigma$ that contains two magnetic events by combining pieces as shown. 
A neutral electric object (dark) comes in and ends on the magnetic event A.  The magnetic event A,  in turn, emits a Dirac membrane (shaded) that is next absorbed by event B, which finally proceeds to emit a charged electric object (dark).  }
 \end{center}
\end{figure}

When proceeding from event A to event B, the Dirac string goes up in time while turning, describing a sort of smooth staircase, i.e., a slide.  The first stage, event A, is the $(x^3,x^4)$-plane at $x^0=x^1=x^2 = 0$ and the second stage, event B, is the $(x^1,x^2)$-plane at $x^0 = a, x^3=0, x^4 = 0$.  

We will take again retarded boundary conditions.  The retarded field $F^{ret}$ is given by the pure Maxwell solution (\ref{FA}) up to $B$ but takes a different form to the future of it.  This is because the non-linear Chern-Simons terms come in for $x^0 > a$.  The calculation that follows will not involve the field after event $B$.  

The total electric charge emitted by the composite magnetic event is the sum of the those emitted by $A$ and  $B$. Since the fields are retarded, the object incoming from the past into event $A$ is neutral, and therefore that event does not emit (or rather, absorb) any charge.  
On the other hand, the event $B$ is `illuminated" by the retarded field of A and, furthermore, due to the turning of the staircase, the illumination induces the emission of an electric charge
\begin{equation}
\label{QB}
Q_B =     \a g \int_{x^0=a} dx^1 dx^2 F^A_{12}.
\end{equation}

Now, if we take for $F^A_{12}$ the expression given by (\ref{FA}) we find that the integral diverges. However, these divergency is spurious and it is actually due to the point character of the source in the present effective theory. Indeed, the field (\ref{FA}) obeys, in the three dimensional space of $(x^0,x^1, x^2)$,
\begin{equation}\label{dd}
{}^* dF=  g\delta^{(3)}(x).
\end{equation}
Therefore, if we close the plane $x^0=a$ by adding to it a timelike  cylinder at infinity, and another plane parallel to $B$ at negative $x^0$, we find, by applying Stokes' formula to (\ref{dd}),  
\begin{equation}
\label{QBB}
Q_B =    \a g^2.
\end{equation}
Since the field is retarded,  the cylinder and the lower plane do not contribute to the integral.  

Note that the result is independent of the time separation $a>0$ between events $A$ and $B$, as one could have anticipated on dimensional grounds. This is an evidence of the topological nature of the charge emission process, which is elaborated upon in the next subsection.

The point, of course, is that in order to apply Stokes'  formula, one needs to regularize the field $F_{\mu\nu}$ on the lightcone, for example, in the manner illustrated in Fig. 3. However, the result (\ref{QBB}) should be independent of the details of the regularization, because any smearing of the delta-function source in (\ref{dd})  (which should be automatically implemented in an underlying non-effective theory)   would preserve Gauss' law. The regularization of the charge density is illustrated in Fig 3.

\begin{figure}[h]
\label{fig2} 
\begin{center}
  \includegraphics[width=8cm]{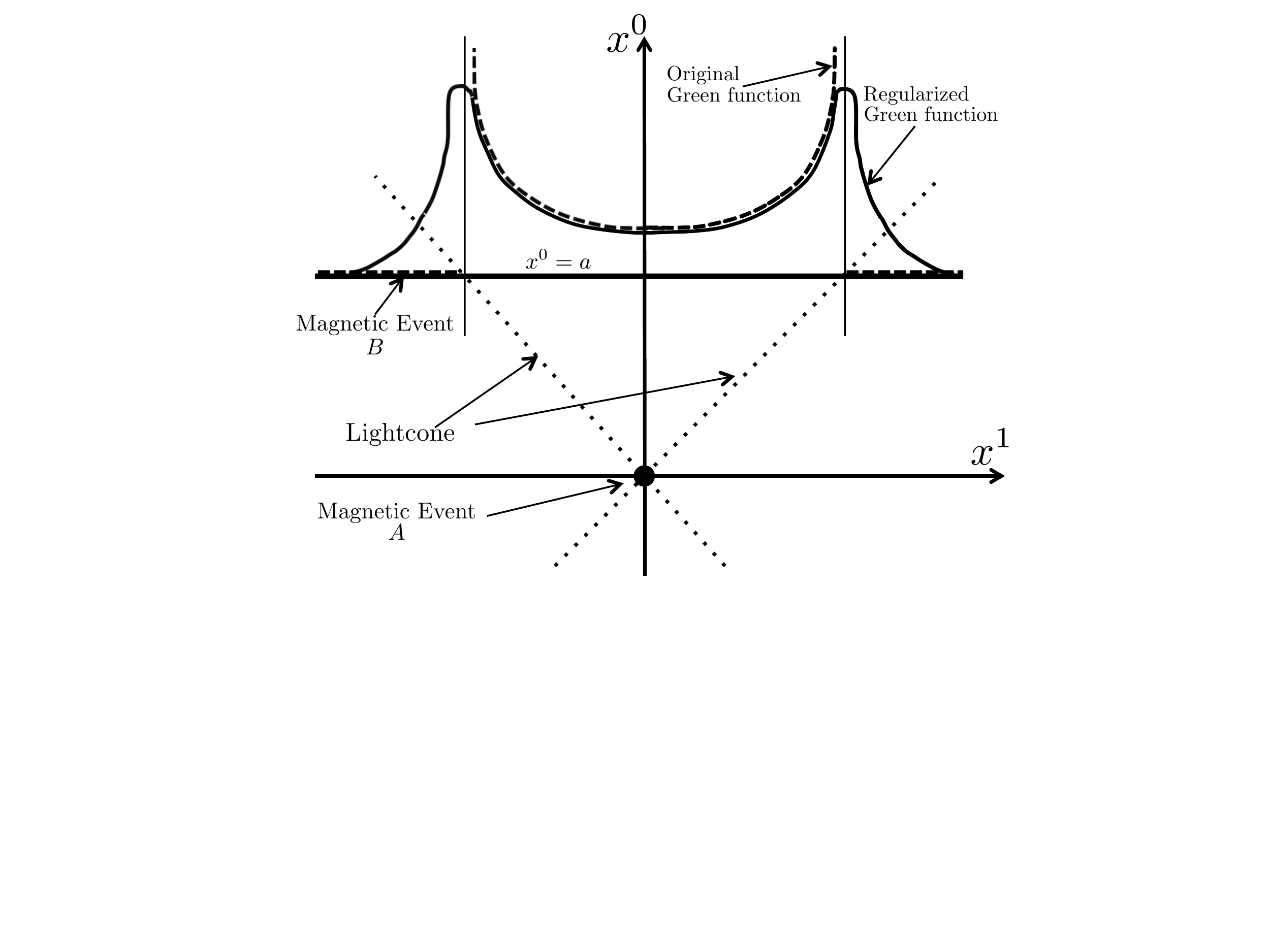}
  \caption{Regularization of the retarded field. The figure depicts the magnetic events $A$ and $B$, located at $x^0=x^1=x^2=0$ and $x^0=a,x^3=x^4=0$ respectively. A section along the $(x^0,x^1)$ plane at $x^3=x^4=0$ is shown. The event $A$ is a point at the origin of this section. The profiles of the retarded field $F_{12}$ on the event B before and after regularization. The original field is non-vanishing inside its future lightcone only, but  diverges on it. After regularization, which may be thought of as the smearing of event $A$, the field becomes finite on the lightcone, but has a rapidly decaying tail outside of it.}
  
     \end{center}
\end{figure}

If we apply the general quantization rule (\ref{eg}) for the product of the electric and magnetic charges to $Q_B$ given by (\ref{QBB}) we conclude that the Chern-Simons coupling must be quantized according to
\begin{equation}
\label{AlphaD=5}
\  \a g^3 =  2 \pi m \hbar \hspace{0.5cm} (D=5)
\end{equation}
This result was previously obtained in \cite{Bachas:1998rg} using dimensional reduction and the Witten effect in four spacetime dimensions, namely  the acquisition of electric charge in the presence of a topological $\theta$-term. 

It is important to realize that the quantization rule for the product of electric and magnetic charges needs that the Dirac string should be able to trace a surface which includes all the electric charge, which will be subject to quantization, without violating the Dirac veto. In the case at hand, although the electric object is infinitely extended in space, and cannot be enclosed by a three dimensional surface, the density of electric charge is effectively confined to a disk (2-ball) within the lightcone of event $A$, as illustrated in Fig. 3. Therefore, provided that the Dirac string remains outside of that ball, the standard quantization argument applies straightforwardly.  

\subsection{Emitted charge as Chern class}

The electric charge emitted by the composite magnetic event admits a simple geometrical description. As it was pointed out in the previous subsection, in order to regularize the emitted charge, it was natural to bring in a closed surface that completed the last magnetic event so as to  link the first one.  Actually, since the self-field of the second event does not contribute to the integral over it, one may take the upper face of the closed surface to lie below the second event. This brings out more clearly the fact that the key geometrical aspect is to link the first event. Thus, we might rewrite (\ref{QB}),
\begin{equation}
\label{emitted}
Q_{emitted}= Q_B =  \a g\oint _{ \sigma^2_A} F, 
\end{equation}
wher $ \sigma^2_A$ is a closed surface that links event A.

The form $F$ is closed on $\sigma^2_A$. This is because there are no magnetic events on $ \sigma^2_A$, except event $B$, which is tangent to it  so that the projection of Eq. (\ref{monopoles}) on $\sigma^2_A$ reduces to $dF=0$. Therefore, one may arbitrarily deform $\sigma^2_A$ below $B$ provided it still links $A$.

Although closed, the $2$-form $F$ is not exact on $ \sigma^2_A$  because its integral over it is different from zero as shown by (\ref{QBB}). As a result of the presence of the magnetic events, the connection $A$ is  defined through local patches on $\sigma^2_A$. If the events are properly chosen so as to generate the complete $D=5$ spacetime, a global connection has a non-vanishing Chern class, which is related to the emitted charge by (\ref{emitted}) and is quantized according to (\ref{QBB}).

We have used, for simplicity, the retarded field for each individual event. If one where to consider the time-reversed situation, the retarded field would be replaced by the advanced field and the charged membrane would be emitted towards the past out of event $A$.  One would then have 
\begin{equation}
\label{emittedA}
Q_{emitted}= Q_A = - \a g\oint _{ \sigma^2_B} F,
\end{equation}
where $\sigma^2_B$ now links event $B$.

The question naturally arises as to whether one could write $Q_{emitted}$ as an integral over a surface $\sigma^2$ which would not have to be chosen according to the boundary condition for the field. We differ this issue to future investigations.

One is familiar with the change in topology in non-abelian gauge theories, when an instanton (solution of the Euclidean field equations) is present in between the remote past and the distant future. In that case the topology change is due to quantum mechanical tunneling. Here, we may also view the charge emission process as one that changes the topology in the sense that the winding number $\int F = \oint_{\infty} A $ changes.  However, in contradistinction with the instanton case, the change in topology is due to the presence of a (composite) magnetic event in Lorenzian spacetime.  It is not connected to quantum mechanical tunneling but already goes classically.

\section{Higher $D$ and $p$}
\label{HigherDp}
\subsection{$p=1$ for any (odd) $D$}
\label{p=1D}

The image of a slide going up in time and turning that arose for $D=5$ suggests an obvious generalization for higher dimensions. 

{}For $D=7$, in which case the magnetic events are four-dimensional, one adds a third elementary event C, which emits a physical object with electric charge $Q \not=0$. Just as before the retarded solution  is exact all the way from $-\infty$ to $t_C$. 

While the turning staircase goes up, the following steps, illustrated in Fig. 4,  occur: (i) A Dirac membrane (shaded) comes in from the past and it is absorbed by a magnetic event A with charge $-g$. (ii) A physical membrane is emitted by A.  This intermediate physical object is  neutral, due to the lack of sufficient illumination.  It moves towards the future as a free Nambu membrane (the ``Lorentz force" on the right side of (\ref{eo}) vanishes).  (iii) Another neutral physical membrane comes in from the past of event B, which has charge $g$.   Note that there is no danger of these neutral membranes to violate the Dirac veto because they do not carry an electric current.    (iv) A Dirac membrane (shaded) is emitted by B and absorded by C, of charge $-g$. (v) A physical object (dark) is emitted by C.

It would have been nice to take the physical membrane absorbed by B to be the one emitted by A. This would have been permissible for a Dirac string which is physically unobservable, but it cannot be done for a physical membrane because it would have to move faster than light at sufficiently large distances.

\begin{figure}[h]
\label{fig5} 
\begin{center}
  \includegraphics[width=8cm]{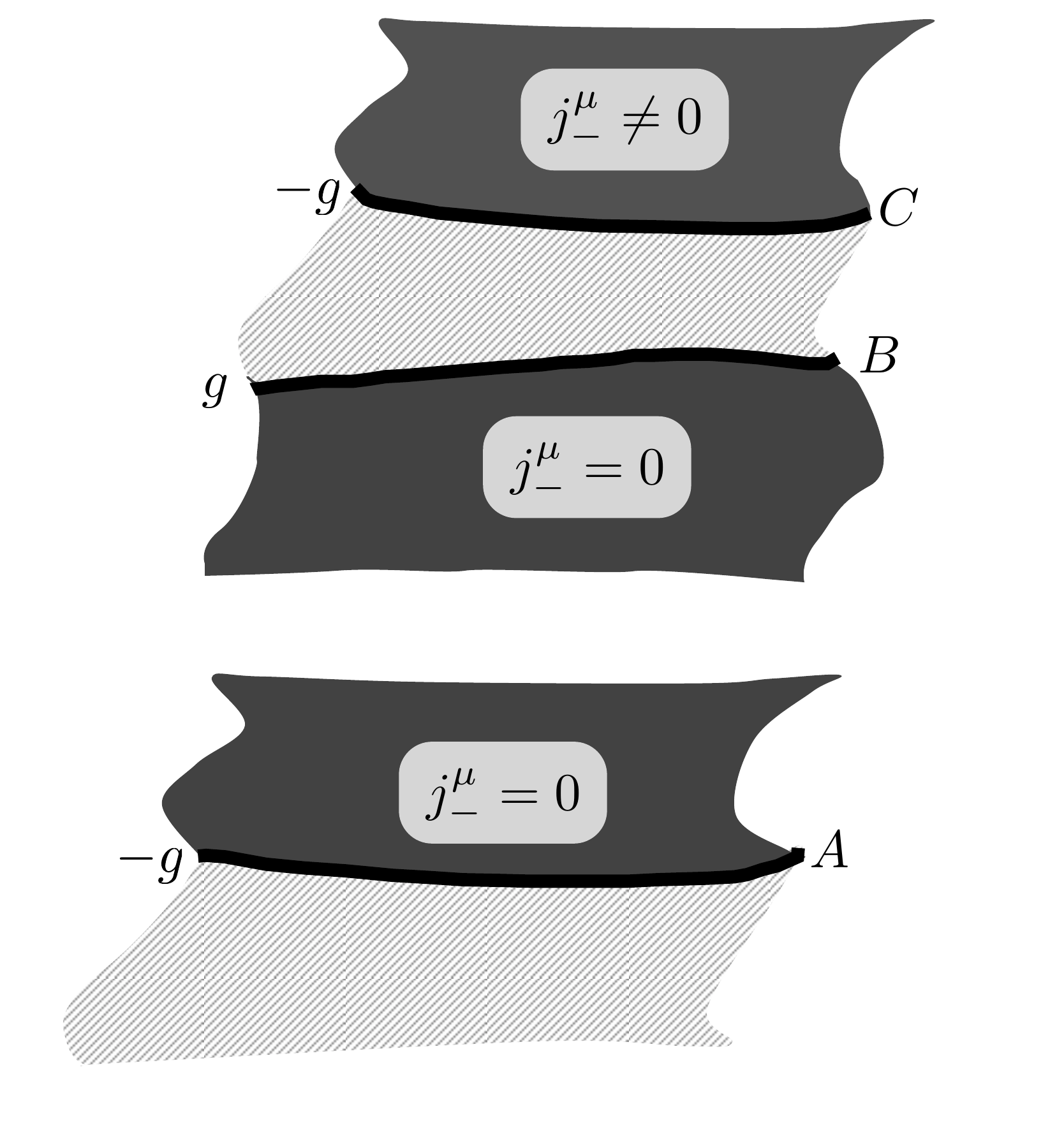}
  \caption{The surface $\Sigma$ for $D=7$.  It consists of two disconnected pieces, one of the type shown in Fig. 1 for a single magnetic event, and another of the type shown in Fig. 2 for a sequence of two magnetic events.  The two pieces cannot be joined to form a single connected surface because the neutral electric object emitted by A cannot be absorbed by B without exceeding the speed of light. This is why a void is left in the figure between the physical neutral membranes emitted by A and absorbed by B, respectively.  That  problem does not arise for Dirac membranes which are physically unobservable. Hence, one can safely connect B and C as shown.}
  \end{center}
\end{figure}

Next one observes that again it is only the last event C that is sufficiently illuminated (by both A and B) to have a non-vanishing emitted charge $Q$.  

One finds for the charge emitted at the last event C the value,   
\begin{equation} \label{QQ}
Q_{emitted} =  \frac{\a g}{2} \int_C F \wedge F  \; \; \; \; (D=7),
\end{equation}
which is the analog of (\ref{Q}).

This integral breaks up as a product
\begin{equation}
\label{QQ2} 
Q_{ emitted} =  \a g\int dx^1 dx^2  F^A_{12} \int dx^3 dx^4 F^B_{34}.
\end{equation}
Each of the two-dimensional integrals appearing in (\ref{QQ2}) is of the type already discussed for $D=5$, the first one has the value $-g$, the second one has the value $+g$. Hence, the emitted charge $Q_{emitted}$ is given by 
\begin{equation} \label{QQ3}
Q_{emitted} =   -  \a g^3 .
\end{equation}

It is also possible in this case to express the emitted charge by replacing event C as the region of integration in (\ref{QQ}) by a closed surface $\sigma_4$ lying below it.  The closed surface $\sigma_4$ maybe constructed as follows. If one imagines events A,B and C as parts of a staircase spiriling upwards, one takes another staircase which turns, say, slightly faster.  This new staircase is taken to start earlier than A and to end before C.  The intersections of events A and B with the initial staircase are four-dimensional. In contradistinction, their intersections with the faster turning staircase are ``generic", i.e. are connected two-dimensional infinite surfaces. The surface $\sigma_4$ is the boundary of the faster turning staircase, including the boundary at infinity.

Just as before, the piece at infinity is not sufficiently illuminated and it does not contribute to the integral. Neither does the bottom plane since the fields are retarded.  Thus one can rewrite (\ref{QQ}) as
\begin{equation} \label{QQbis}
Q_{emitted} =  \frac{\a g}{2} \oint_{\sigma_4} F \wedge F  \; \; \; \; (D=7),
\end{equation}
For $D=5$, the intersection of event A with the corresponding faster turning staircase, which can be continuously  deformed to any three-dimensional surface bounded by the surface $\sigma_2$ used there, is a point, and therefore the linking number appears. For $D=7$, what appears is an interesting extension of the latter notion.

Note that by using Stokes theorem and (\ref{monopoles}),  one may go down the staircase, rewrite (\ref{QQ}) as,
\begin{equation}
\label{emitted7}
Q_{emitted}=  \alpha g^2 \left( \oint _{ \sigma^2_A} F\right),
\end{equation}
and appeal to (\ref{emitted}) to recover (\ref{QQ3}).

Demanding that $Q$ be quantized according to Dirac's rule, yields the quantization of the Chern-Simons coupling
\begin{equation}\label{d7qc}
 \alpha g^4 = 2\pi \hbar m  \ \ \ (D=7)
\end{equation}

{}For $p=1$ and $D = 2n+1$,   the staircase has dimension $2n -1$ and contains $n$ magnetic events $E_i$ of dimension $2n-2$ and magnetic charges of alternating signs. The last event $E_n$ has magnetic charge $-g$.

One finds that only the last event $E_n$ is sufficiently illuminated and emits a charge,
\begin{equation}
Q_{emitted} =   \frac{\alpha g}{(n-1)!}\int _{E_n} \underbrace{F\wedge\cdots\wedge F}_{n-1 \mbox{ \small times}} =  \; \a \, g^{n} .
\end{equation}
The Chern-Simons coupling $\a$ is then quantized according to
\begin{equation}
  \a \, g^{n} = 2\pi \hbar m \ \ \ (D=2n+1)
\end{equation}

We have normalized  $\a$ for any $n$ through, 
\begin{eqnarray}\label{CS'}
I_{CS} &=& \frac{\alpha}{(n+1)!}\int \underbrace{dA\wedge \cdots \wedge dA }_{n \mbox{ \small times}}\wedge A  + \nonumber \\ 
&+& \frac{\alpha}{n!}\int {}^*G\wedge \underbrace{dA \wedge \cdots \wedge dA}_{n-1 \mbox{ \small times}}\wedge A .
\end{eqnarray}

\subsection{$p=3$, $D=11$}

The procedure can be applied straightforwardly if one replaces the basic $1$-form $A$ by a $p$-form (with $p$ odd) and builds the corresponding Chern-Simons forms out of it.  For the existence of the Chern-Simons forms, one must have $D = kp+k + p$ for some integer $k$. In this case, the role that charged particles played for $p=1$ is played by extended objects of worldvolume dimension $p$, while the corresponding dimensions of the magnetic event and the emitted electric object are $D-p-2$ and $D-p-1$, respectively \cite{Teitelboim:1985yc}.  

In particular for the interesting case $p=3$, $D=11$, which arises in supergravity, the Chern-Simons action reads
\begin{equation}\label{CS''}
I_{CS}=  \frac{\alpha}{6}\int dA\wedge dA \wedge A + \frac{\alpha g}{2}\int_{\Sigma^7}  dA \wedge A .
\end{equation}
which is of the same form as the one for $p=1$, $D=5$ and therefore the coupling is now quantized as
\begin{equation}
\a g^3 = 2\pi \hbar m.
\end{equation}

Since for p=3, $D=11$, the point particles of $p=1$, $D=5$ are replaced by two dimensional extended objects, the analog of the composite event discussed in  subsection  IV B would consist of two  six-dimensional events  separated by a three-dimensional slab (instead of the one dimensional time axis interval). Thus, one could take event $A$ to have the equations  $x^0=0$;  $x^3=x^4=x^5=x^6=0$ and event B to be defined by $x^0=a$; $x^7=x^8=x^9=x^{10}=0$. The joining slab would then have $0<x^0<a$, 
$-\infty< x^1,x^2<+\infty$.  

For $p \not=1$, athough there is an underlying $U(1)$ gauge invariance, the corresponding electric and magnetic charges are not those of ordinary Maxwell theory. The conserved electric current is not the dual of $(D-1)$-form but rather the dual of a $(D-p)$-form and the charge is computed by integration of the dual of the current over a $(D-p)$ submanifold  \cite{Teitelboim:1985ya}.

\section{Concluding remarks}

Recently, the concept of extended charged events has been introduced, and it has been argued, on the basis of the principle of electric-magnetic duality,  that events should play as central a role as that played by particles or ordinary branes. In this article we show that in the presence of a Chern-Simons coupling, a magnetically charged extended event emits an extended physical object which may be electrically charged, and we write an action principle which accounts for this effect. The action involves two Chern-Simons terms, one integrated over spacetime, and another integrated over the submanifold obtained by joining the history of the emitted physical object and the worldsheet of the Dirac string of the magnetic event. 

By demanding that the total emitted charge of a composite extended magnetic event be quantized according to Dirac's rule, we find a quantization condition for the Chern-Simons coupling.  For a $1$-form electric potential in $D=2n+1$ spacetime dimensions, the composite object is formed by $n$ extended magnetic events separated in time  such that the product of their transverse  spaces, together with the time axis, is the entire spacetime. 

Our procedure is a generalization of the one used in Ref. \cite{Henneaux:1986tt} in $D=3$. However, in $D=5$ and higher, the problem is much richer geometrically and presents, for example, the novel phenomena of induced electric charge emission. 

For the case $D=11$, $p=3$, our result for the quantization of the Chern-Simons coupling was obtained previously in \cite{Duff:1995wd,deAlwis:1996ez,deAlwis:1997gq,Witten:1996md,Bachas:1998rg} in the context of M-theory (i.e. through the use of a combination of results from 11-dimensional supergravity and string theory).  One cannot help feeling that this agreement is remarkable and makes the existence of events even more compelling since M-theory is ready to accomodate them.

\acknowledgements
We would like to thank Constantin Bachas for many illuminating discussions on charged events, and Cristi\'an Mart\'{\i}nez for much help in the preparation of the manuscript.
M. H. gratefully acknowledges support from the Alexander von Humboldt Foundation through a Humboldt Research Award and support from the ERC through the ``SyDuGraM" Advanced Grant. The Centro de Estudios Cient\'{\i}ficos (CECs) is funded by the Chilean Government through the Centers of Excellence Base Financing Program of Conicyt. The work of M. H. is also partially supported by IISN
- Belgium (conventions 4.4511.06 and 4.4514.08), by the Belgian Federal Science Policy Office through the Interuniversity Attraction Pole P6/11 and by the ``Communaut\'e Fran\c{c}aise de Belgique" through the ARC program. The work of AG was partially supported by Fondecyt (Chile) Grant \#1090753.

\end{document}